\documentclass[superscriptaddress,nofootinbib,showkeys,showpacs,byrevtex,preprintnumbers]{revtex4}
\usepackage{graphicx}
\usepackage{epstopdf}
\begin{document}
\preprint{PKNU-NuHaTh-2017-02}
\title{Quasi-distribution amplitudes for pion and kaon via the nonlocal chiral-quark model}
\author{Seung-il Nam}
\email[E-mail: ]{sinam@pknu.ac.kr}
\affiliation{Department of Physics, Pukyong National University (PKNU), Busan 608-737,
Republic of Korea}
\affiliation{Asia Pacific Center for Theoretical Physics (APCTP)f, Pohang 790-784,
Republic of Korea}
\date{\today}
\begin{abstract}
We investigate the pseudoscalar (PS) meson ($\pi$ and $K$) quasi-distribution amplitude (QDA), which is supposed to be an asymptotic analog to the meson distribution amplitude (DA) $\phi_{\pi,K}(x)$ in the limit of the large longitudinal PS-meson momentum, i.e.  $p_3\to\infty$, in the nonperturbative region. For this purpose,  we employ the nonlocal chiral-quark model (NLChQM) in the light-front formalism with a minimal Fock-state for the mesons $\sim q\bar{q}$ at the low-energy scale parameter of the model $\Lambda\sim1$ GeV. As a trial, we extract the transverse-momentum distribution amplitude (TMDA) from the light-front wave function within the model, and convert it to QDA with help of the virtuality-distribution amplitude. By doing that, we derive an analytical expression for the nonperturbative QDA with the current-quark mass correction up to $\mathcal{O}(\Delta m_q)$. Numerically, we confirm that the obtained TMDA reproduces the experimental data for the photon-pion transition form factor $F_{\gamma\gamma^*\pi^0}(Q^2)$ at the low-$Q^2$ qualitatively well. We also observe that the obtained QDA approaches to DA as $p_3$ increases, showing the symmetric and asymmetric curves with respect to $x$ for the pion and kaon, respectively, due to the current-quark mass difference $m_{u,d}\ll m_s$. Assigning $\xi\equiv2x-1$, the moments $\langle\xi^n\rangle_{\pi,K}$ are computed, using the pion and kaon QDAs, and there appear only a few percent deviations in the moments for $p_3\gtrsim30\Lambda$ in comparison to the values calculated directly from DAs. It turns out that the higher moments are more sensitive to the change of $p_3$, whereas the lower ones depend less on it.
\end{abstract}
\maketitle
\section{Introduction}
For decades, a hard (semi-inclusive or exclusive) process of QCD have been explored extensively in terms of the factorization theorem, which separates the process into the process-dependent short- and the universal long-distance parts~\cite{Collins:1989gx,Lepage:1979za,Lepage:1979zb}. The former can be understood by the perturbative calculation of QCD, whereas the latter needs much more complicated nonperturbative knowledge for the hadrons. Among those nonperturbative quantities, the leading-twist (twist-2) distribution amplitudes (DA) for the pseudoscalar (PS) mesons are of importance to describe the exclusive meson-production processes, since it gives essential information to the quark momentum distribution inside the meson~\cite{Lepage:1979zb,Efremov:1979qk,Lepage:1980fj,Chernyak:1983ej}.  Experimentally, the pion DA was explored by investigating the pion-photon transition form factor by the CELLO~\cite{Behrend:1990sr}, CLEO~\cite{Gronberg:1997fj}, BaBar~\cite{Aubert:2009mc}, and Belle~\cite{Uehara:2012ag}. The analysis of those data in Ref.~\cite{Schmedding:1999ap} has suggested that neither two-humped DA predicted by Chernyak and Zhitnitsky (CZ)~\cite{Chernyak:1977fk} nor the asymptotic (AS) one are favored at the $2\sigma$ level of accuracy. These observations were supported by Bakulev {\it et al.} in
Refs.~\cite{Bakulev:2002uc,Bakulev:2003cs}.  From the theoretical studies, DA has been investigated in various
approaches: In the QCD sum rules (QCDSR)~\cite{Chernyak:1983ej,Braun:1988qv,Bakulev:1991ps,Bakulev:1994su,Bakulev:2001pa,Bakulev:2005vw}, in lattice QCD (LQCD)~\cite{Dalley:2002nj}, in the nonlocal chiral-quark model (NLChQM) from the instanton vacuum~\cite{Petrov:1998kg,Nam:2006au,Nam:2006sx}, in the NJL models~\cite{Praszalowicz:2001wy,Praszalowicz:2001pi,RuizArriola:2002bp}, and so on.

In general, the PS-meson DA itself has been computed in the light-front (LF) formalism, which is not a covariant one, to consider the well-defined Fock states of the meson with the Bjorken variable $x$. Hence, the lower moments of DA have been computed in LQCD simulations, which are performed in Euclidean space, instead of computing DA directly. In Ref.~\cite{Ji:2013dva}, the author proposed an asymptotic analog of DA in the limit of the large longitudinal momentum of the meson $p_3$, i.e. quasi distribution amplitude (QDA), which is the matrix elements of equal-time bilocal operators as a function of $y=[-\infty,\infty]\in\textbf{R}$ and $p_3$. The basic idea of Ref.~\cite{Ji:2013dva} is to connect the LF correlation for the partons to the equal-time correlation  in the infinite-momentum frame. Related works using the idea have been carried out via various approaches including LQCD simulations as well~\cite{Zhang:2017bzy,Carlson:2017gpk}. In a model calculation of Ref.~\cite{Radyushkin:2017gjd}, the transverse-momentum distribution amplitude (TMDA) was proposed in a space-like separation of the partons as a function of the transverse momentum $k^2_\perp$ and $x$. After defining the virtuality distribution amplitude (VDA) with QDA at the equal-time bilocal operator, a relation between TMDA and QDA was settled. Using some model parameterizations for TMDA, such as the CZ and Asymptotic DAs with the Bessel and Gaussian functions of $k^2_\perp$, QDA was explored with different $p_3$ values. It was turned out that the nonperturbative evolution stops at $p_3/\Lambda\sim20$ GeV, whereas the perturbative evolution ends up with the asymptotic DA $\phi_\mathrm{asym}=6x(1-x)$. For the perturbative evolution $p_3\gtrsim20$ GeV, the one-gluon exchange loop diagrams were also taken into account~\cite{Radyushkin:2015gpa}.

In the present work, we employ NLChQM to study QDA at the low renormalization scale $\Lambda\sim1$ GeV. The nonperturbative TMDA is derived from the light-front wave function (LFWF) from the model calculation in the nonperturbative region as a trial to solve the integral equation, which relates TMDA and LFWF. Then, following the procedure suggested in Ref.~\cite{Radyushkin:2017gjd}, we construct QDA as a function of $p_3$ and $y$. All the model parameters are determined to reproduce the PS-meson weak-decay constants and to satisfy the normalization condition for DAs. We obtain an analytic expression for QDA for the PS-meson QDA in terms of the quark mass difference $\Delta M=|M_q-M_{\bar{q}}|\propto\Delta m_q$. We verified numerically that QDA approaches to DA as $p_3$ increases as expected, and it shows the symmetric and asymmetric shapes for the pion and kaon DAs, due to the quark-mass differences. We also observe that, for $p_3\gtrsim80$ GeV, the difference between the QDA and DA becomes almost negligible. The moments $\langle\xi^n\rangle_{\pi,K}$ are also computed using DA as well as QDA. It turns out that the higher moments depend much on the change of $p_3$, while the lower ones less on it. At $p_3=30\Lambda\approx30$ GeV, the moments show only a few percent differences, when those from DA and QDA are compared.

The present manuscript is organized as follows: In Section II, we briefly introduce VDA, TMDA, and QDA, then make explanations for TMDA and LFWF in the nonperturbative regions. The effective nonperturbative model, i.e. NLChQM is explained, and an analytical expression for QDA is derived from the model in Section III. The numerical results for QDA and the moments, and corresponding explanations are given in Section IV. Section V is devoted to conclusion and summary.
 \section{Relations between the distribution amplitudes}
In this Section, we make a brief explanation for the theoretical derivation of QDA, closely following Ref.~\cite{Radyushkin:2017gjd}. After doing that, we suggest a relation between TMDA and LFWF in the nonperturbative region. The matrix element for an axial-vector current can be written in terms of VDA $\Phi_\mathcal{M}(x,\sigma)$ for the PS mesons $(\mathcal{M})$:
\begin{equation}
\label{eq:VDA}
\langle0|\bar{q}(0)\gamma_5\gamma_\mu q(z_3)|\mathcal{M}(p)\rangle=\frac{1}{\sqrt{2}F_\mathcal{M}}\int^\infty_0d\sigma\int^1_0dx
\left[p_\mu\Phi_\mathcal{M}(x,\sigma)+z_\mu Z_\mathcal{M}(x,\sigma)\right]
e^{-ix(p\cdot z)-i\sigma z^2/4},
\end{equation}
where $p$ and $\sigma$ stand for the PS-meson four momentum and the virtuality-like parameter $\propto\mathrm{mass}^2$, respectively. $F_\mathcal{M}$ denotes the weak-decay constant for the mesons and we choose $F_{\pi,K}=(93,113)$ MeV empirically. Taking the light front $z_+=0$ with $z_\perp=0$, the leading-twist contribution with the definition $p=(p_0,0,0,p_3)$ reads:
\begin{equation}
\label{eq:VDAL}
\langle0|\bar{q}(0)\gamma_5\gamma_+ q(z_-)|\mathcal{M}(p)\rangle=
\frac{p_+}{\sqrt{2}F_\mathcal{M}}\int^\infty_0d\sigma\int^1_0dx\,
\Phi_\mathcal{M}(x,\sigma)e^{-ixp_+z_-}.
\end{equation}
Hence, the twist-2 DA $\phi(x)$ and VDA satisfy the following relation and DA is normalized as:
\begin{equation}
\label{eq:DAVDA}
\phi_\mathcal{M}(x)=\int^\infty_0d\sigma\,\Phi_\mathcal{M}(x,\sigma),\,\,\,\,\int^1_0dx\,\phi_\mathcal{M}(x)=1
\end{equation}
in the definition of Eq.~(\ref{eq:VDAL}). We want to introduce TMDA $\Psi(x,k^2_\perp)$ as a Fourier transformation for the matrix element in the l.h.s. of Eq.~(\ref{eq:VDAL}) with respect to $k_\perp$:
\begin{equation}
\label{eq:TMDA}
\langle0|\bar{q}(0)\gamma_5\gamma_+ q(z_-)|\mathcal{M}(p)\rangle
=\frac{p_+}{\sqrt{2}F_\mathcal{M}}\int^\infty_0d^2k_\perp\int^1_0dx\,\Psi_\mathcal{M}(x,k^2_\perp)\,e^{-ixp_+z_-}.
\end{equation}
Taking into account the relation between DA and VDA in Eq.~(\ref{eq:DAVDA}), one can write TMDA in terms of VDA with a Gaussian factor as a function of $k^2_\perp$ as follows:
\begin{equation}
\label{eq:TMDAVDA}
\Psi_\mathcal{M}(x,k^2_\perp)=\frac{i}{\pi}\int^\infty_0\frac{d\sigma}{\sigma}\,\Phi_\mathcal{M}(x,\sigma)
e^{-ik^2_\perp/\sigma}.
\end{equation}
From the definition of Eq.~(\ref{eq:TMDAVDA}), it is obvious that the integration TMDA over $k_\perp$ gives DA with Eq.~(\ref{eq:DAVDA}), due to the Gaussian integration over $k^2_\perp$:
\begin{equation}
\label{eq:TMDADA}
\phi_\mathcal{M}(x)=\int^\infty_0 dk^2_\perp\,\Psi_\mathcal{M}(x,k^2_\perp)
=2\pi\int^\infty_0 k_\perp dk_\perp\,\Psi_\mathcal{M}(x,k^2_\perp).
\end{equation}
Thus, the relation between TMDA and DA has been settled.

Now, we are in a position to define QDA $Q_\mathcal{M}(y,p_3)$ with the matrix element given in Eq.~(\ref{eq:VDA}) as an equal-time bilocal operator with $z=(0,0,0,z_3)$ and $\mu=0$:
\begin{equation}
\label{eq:VDA2}
\langle0|\bar{q}(0)\gamma_5\gamma_0 q(z_3)|\mathcal{M}(p)\rangle=\frac{p_0}{\sqrt{2}F_\mathcal{M}}
\int^\infty_0d\sigma\int^1_0dx \,\Phi_\mathcal{M}(x,\sigma)e^{-ixp_3 z_3+i\sigma z^2_3/4}.
\end{equation}
Being similar to the procedure for introducing TMDA as the Fourier transformation, we define the following:
\begin{equation}
\label{eq:QDA}
\langle0|\bar{q}(0)\gamma_5\gamma_0 q(z_3)|\mathcal{M}(p)\rangle=\frac{p_0}{\sqrt{2}F_\mathcal{M}}
\int^\infty_{-\infty}dy\,Q_\mathcal{M}(y,p_3)e^{-iyp_3 z_3}.
\end{equation}
It is easy to check that Eq.~(\ref{eq:VDA2}) equates with Eq.~(\ref{eq:QDA}), if we write QDA as follows:
\begin{equation}
\label{eq:QDAVDA}
Q_\mathcal{M}(y,p_3)=\int^1_0dx\int^\infty_0d\sigma\sqrt{\frac{ip^2_3}{\pi\sigma}}
e^{-i(x-y)^2p^2_3/\sigma}\Phi_\mathcal{M}(x,\sigma),
\end{equation}
since the integration QDA over $y=[-\infty,\infty]$ is appropriately normalized according to Eq.~(\ref{eq:DAVDA}). We try to replace $k^2_\perp$ into $k^2_1+(x-y)^2p^2_3$ in Eq.~(\ref{eq:TMDAVDA}), and integrate it over $k_1$ and multiply it by $p_3$, resulting in 
\begin{eqnarray}
\label{eq:QDATMDA222}
p_3\int^\infty_{-\infty}d k_1\,\Psi_\mathcal{M}(x,k^2_1+(x-y)^2p^2_3)
=\int^\infty_0d\sigma\,\sqrt{\frac{ip_3^2}{\pi\sigma}}
e^{-i(x-y)^2p^2_3/\sigma}\Phi_\mathcal{M}(x,\sigma).
\end{eqnarray}
Comparing Eqs.~(\ref{eq:QDAVDA}) and (\ref{eq:QDATMDA222}), we finally obtain the following relation between QDA and TMDA:
\begin{equation}
\label{eq:QDATMDA}
Q_\mathcal{M}(y,p_3)=p_3\int^\infty_{-\infty}dk_1\int^1_0dx\,\Psi_\mathcal{M}\left(x,k^2_1+(x-y)^2p^2_3\right).
\end{equation}
Note that QDA in Eq.~(\ref{eq:QDAVDA}) is derived in a covariant frame. From now on, we want to investigate the properties of QDA in detail. The integrand excluding $\Phi_\mathcal{M}(x,\sigma)$ in the r.h.s. of Eq.~(\ref{eq:QDAVDA}) satisfies 
\begin{equation}
\label{eq:LIM}
\lim_{p_3\to\infty}\left[\sqrt{\frac{ip^2_3}{\pi\sigma}}e^{-i(x-y)^2p^2_3/\sigma}\right]=\delta(x-y).
\end{equation}
Using this limit for Eq.~(\ref{eq:QDATMDA}), we have the following 
\begin{equation}
\label{eq:QDAVDA2}
\lim_{p_3\to\infty}Q_\mathcal{M}(y,p_3)\equiv \varphi_\mathcal{M}(y)
=\int^1_0dx\int^\infty_0d\sigma\,\delta(x-y)\Phi_\mathcal{M}(x,\sigma),
=\int^\infty_0d\sigma\,\Phi_\mathcal{M}(y,\sigma).
\end{equation}
Thus, by taking the limit $p_3\to\infty$ for QDA, one can obtain a DA-like distribution $\varphi_\mathcal{M}(y)$ for $y=[-\infty,\infty]$ in a covariant manner. In this way, from LQCD simulations for QDA, the information for DA can be extracted by matching DA and QDA~\cite{Radyushkin:2017gjd,Ji:2013dva}. It is also easy to see that QDA for $p_3\to\infty$, i.e., $\varphi_\mathcal{M}(y)$ and DA satisfy the normalization conditions, because of Eq.~(\ref{eq:DAVDA}):
\begin{equation}
\label{eq:NORM2}
\int^\infty_{-\infty}dy\,\varphi_\mathcal{M}(y)=\int^1_0dx\,\phi_\mathcal{M}(x)=1.
\end{equation}

Considering Eqs.~(\ref{eq:QDATMDA}) and (\ref{eq:NORM2}), which are frame-independent by definition, we can write
\begin{equation}
\label{eq:MOMENT22}
\lim_{p_3\to\infty}p_3\int^\infty_{-\infty}dy\,\int^\infty_{-\infty}dk_1\,
\Psi_\mathcal{M}(x,k^2_1+(x-y)^2p^2_3)=\int^\infty_0d^2k_\perp\,
\psi_\mathcal{M}(x,k^2_\perp).
\end{equation}
It is nontrivial to solve the integral equation in Eq.~(\ref{eq:MOMENT22}), because LFWF $\psi_\mathcal{M}(x,k^2_\perp)$ and TMDA $\Psi_\mathcal{M}(x,k^2_\perp)$ may have different functional forms in principle. However, taking into account the momentum replacement $k^2_\perp\to k^2_1+(x-y)^2p^2_3$ as described above, it is rather rational to introduce a Dirac delta function to solve the integral equation of Eq.~(\ref{eq:MOMENT22}) as follows:
\begin{equation}
\label{eq:MOMENT3}
\lim_{p_3\to\infty}p_3\int^\infty_{-\infty}dy\int^\infty_{-\infty}dk_1\,\int^\infty_{-\infty}dk_2\,
\Psi_\mathcal{M}(x,k^2_\perp)\,\delta\left(k_2-(x-y)p_3\right)
=\int^\infty_0d^2k_\perp\,\psi_\mathcal{M}(x,k^2_\perp)
\end{equation}
with the momentum integral variable $k_2$, which satisfies $k^2_1+k_2^2=k^2_\perp$. Integrating the l.h.s. of Eq.~(\ref{eq:MOMENT3}) with respect to $y$ and considering $\int dk_1dk_2=\int d^2k_\perp$, we arrive at
\begin{eqnarray}
\label{eq:MOMENT4}
\lim_{p_3\to\infty}\Psi_\mathcal{M}\left(x,k^2_\perp\right)\Big|_{x=y+\frac{k_2}{p_3}}=\psi_\mathcal{M}(x,k^2_\perp).
\end{eqnarray}
If $k_2$ is small and finite, i.e., $k_2\sim\Lambda$, where $\Lambda$ denotes a nonperturbative (NP) scale parameter, we obtain the following equation with $\lim_{p_3\to\infty}k_2/p_3=0$:
\begin{eqnarray}
\label{eq:MOMENT5}
\Psi^\mathrm{NP}_\mathcal{M}(y,k^2_\perp)=\psi^\mathrm{NP}_\mathcal{M}(y,k^2_\perp)\,\,\,\,\mathrm{for}\,\,\,\,
y=x=[0,1].
\end{eqnarray}
Therefore, we can conclude that, in the nonperturbative regions, LFWF becomes equivalent to TMDA in the limit $p_3\to\infty$. A similar discussion for the nonperturbative properties of the integrated TMDA was given in Ref.~\cite{Radyushkin:2017gjd} as well. Additionally, in order to test the validity of Eq.~(\ref{eq:MOMENT5}) phenomenologically, in Section IV, we will explicitly compute the pion-photon transition form factor $F_{\gamma\gamma^*\pi_0}$, which is defined by the integration of TMDA over $x$ and transverse momenta as in Eq.~(\ref{eq:FF}), and compare it with experimental data. As will be seen in Fig.~\ref{FIGFF}, Eq.~(\ref{eq:MOMENT5}) works relatively well.

Hence, our strategy is as follows: Although LFWF is not derived in a covariant frame and the solution for Eq.~(\ref{eq:MOMENT22}) is \textit{not unique}, we want to extract QDA from LFWF $\sim$ TMDA in the nonperturbative regions as discussed above, and LFWF will be obtained from an effective nonperturbative model in the next Section.

\section{Pion and kaon distribution amplitudes from NLChQM}
In this Section we make a brief introduction of an effective model for the nonperturbative QCD, based on the liquid-instanton model (LIM), i.e. nonlocal chiral-quark model (NLChQM)~\cite{Diakonov:2002fq}. In LIM, the (anti)instanton as a nonperturbative gluon and semi-classical solution of the Yang-Mills equation plays an important role. For instance, the nontrivial interaction between the instantons and quarks produces the dynamically-generated effective quark mass, which is the signal of the spontaneous chiral-symmetry breaking (S$\chi$SB)~\cite{Diakonov:2002fq}. In addition, since the interaction is nonlocal, the effective quark mass depends on the momentum virtuality in terms of the quark zero-mode solution in the instanton ensemble and performs the role of a UV regulator by construction. Hence, the loop divergences appearing in correlation matrix elements are tamed, being similar to the inclusion of form factors by hand in usual effective QCD-like models, such as the (local) Nambu--Jona-Lasinio (NJL) model. In Euclidean space, the effective partition function of the model for the flavor SU($N_f$) reads~\cite{Diakonov:2002fq}:
\begin{eqnarray}
\label{eq:EZ}
\mathcal{Z}_\mathrm{eff}[q,q^\dagger]&=&\int\frac{d\lambda_\pm}{2\pi}\int Dq Dq^\dagger
\exp\left[
\int_x\sum_{q}\psi_q^\dagger(i\rlap{/}{\partial}+m_q)\psi_q+\sum_{a=\pm}\left[\lambda_a Y^a_{N_f}(\bar{\rho})+N_a\left(\ln\frac{N_a}{\lambda_a V \mathrm{M}^{N_f}}\right)\right]\right],
\cr
Y^a_{N_f}(\bar{\rho})&=&\frac{1}{N^{N_f}_c}\int_x \,
\mathrm{det}_f\left[iJ^a_{q\bar{q}'}(x,\bar{\rho})\right]=
\int_x\,\mathrm{det}_f\left[\frac{i}{N_c}J^a_{q\bar{q}'}(x,\bar{\rho})\right],
\cr
J^a_{q\bar{q}'}(x,\bar{\rho})&=&\int_k\int_pe^{i(k-p)\cdot x}F(k)F(p)\left[\psi_q^\dagger(k)\frac{1+a\gamma_5}{2}\psi_{q'}(p)\right]_{N_f\times N_f}.
\end{eqnarray}
where $q$ and $q'$ indicate different flavors of the quarks. We assign $\int\frac{d^4k}{(2\pi)^4}$ with $\int_k$ for brevity, while $[\cdots]_{N_f\times N_f}$ and $\mathrm{det}_f$ stands for the $(N_f\times N_f)$ quark-flavor matrix and the determinant over the flavor indices. The parameter $a=\pm1$ indicates the instanton ($+$) and anti-instanton ($-$) contributions, whereas $N_a$ and $\lambda_a$ denote the number of (anti)instantons and the Lagrange multiplier, respectively. $V$ and $\mathrm{M}$ stand for the four-dimensional volume in Euclidean space and a parameter to make the logarithm argument dimensionless, respectively. Note that the value of $\mathrm{M}$  does not make any difference in the numerical calculations. Assuming that the QCD vacuum is CP-invariant, i.e. the numbers of the instanton and anti-instanton are the same $N_+=N_-$, and performing the (semi) bosonization for the flavor-SU(3) sector, one arrives at the following effective low-energy chiral action~\cite{Diakonov:2002fq}:
\begin{equation}
\label{eq:ACTION1}
\mathcal{S}_{\mathrm{eff}}[m_q,\mathcal{M}]=-\mathrm{Sp}
\ln\left[i\rlap{/}{\partial}+im_q+i\sqrt{M_q(\partial^2)}U^{\gamma_5}(\mathcal{M})
\sqrt{M_q(\partial^2)}\right],
\end{equation}
where $m_q$, $\mathcal{M}$, and $\mathrm{Sp}$ indicate the current-quark mass,
the pseudo-Nambu-Goldstone (NG) boson field, and the functional
trace over all relevant spaces, respectively. Assuming isospin
symmetry and explicit flavor SU(3) symmetry breaking, we use the following
numerical values $m_{u}=m_d=5$ MeV and $m_s=135$ MeV. The
$M(i\partial)$ stands for the momentum-dependent effective quark mass,
generated from the quark zero modes of the instantons~\cite{Diakonov:1985eg}.  Its analytical form in
the Euclidean four-momentum $(k_E)$ space is given by
\begin{equation}
\label{eq:MMMD}
M_q(t)=M_qF(t),\,\,\,\,
F(t)=2t\left[I_0(t)K_1(t)-I_1(t)K_0(t)-\frac{1}{t}I_1(t)K_1(t) \right].
\end{equation}
Here, $t=|k_E|\bar{\rho}/2$, and $I_\ell$ and $K_\ell$ stand for the modified Bessel functions with the order $\ell$. Note that $\bar{\rho}$ stands for the average instanton size $\sim1/3$ fm~\cite{Diakonov:2002fq}. In the numerical calculations, instead of using Eq.~(\ref{eq:MMMD}), we will make use of the following parametrization for numerical convenience:
\begin{equation}
\label{eq:EFFM}
M_q(k^2_E)=M_qF(k^2_E)=(m_q+M_{0})
\left(\frac{\Lambda^2}{\Lambda^2+k^2_E} \right)^{2},
\end{equation}
where $M_{0}$ indicates the constituent-quark mass in the chiral limit. The pseudo-NG boson field is represented in a
nonlinear form as~\cite{Diakonov:1995qy}:
\begin{equation}
\label{eq:CHIRALFIELD}
U^{\gamma_5}(\mathcal{M})=
\exp\left[\frac{i\gamma_{5}(\lambda\cdot\mathcal{M})}{F_\mathcal{M}}\right]
=1+\frac{i\gamma_{5}(\lambda\cdot\mathcal{M})}{F_\mathcal{M}}
-\frac{(\lambda\cdot\mathcal{M})^{2}}{2F^{2}_\mathcal{M}}+\cdots,
\end{equation}
where $\mathcal{M}^{\alpha}$ is the flavor SU(3) multiplet defined as
\begin{equation}
\label{eq:PHI}
\lambda\cdot\mathcal{M}=\left(
\begin{array}{ccc}
\frac{\pi^{0}}{\sqrt{2}}+\frac{\eta^2}{\sqrt{6}}&\pi^{+}&K^+\\
\pi^-&-\frac{\pi^{0}}{\sqrt{2}}+\frac{\eta^2}{\sqrt{6}}&K^0\\
K^-&\bar{K}^0&-\frac{2\eta^2}{\sqrt{6}}\\
\end{array}
 \right),
 \end{equation}
Here, $F_\mathcal{M}$ denotes the weak-decay constant for the PS mesons, whose empirical values are about $93$ MeV for the pion and $113$ MeV for the kaon for instance. Various matrix elements can be computed within the model by performing the functional differentiation of the effective chiral action with respect to source fields. As for the pion weak-decay constant, by combining the effective chiral action and PCAC, we arrive at the  following expression for the pion weak-decay constant in the chiral limit (CL)~\cite{Nam:2006ng}:
\begin{equation}
\label{eq:FPI}
F_{\pi,\mathrm{CL}}=4N_c\int\frac{d^4k_E}{(2\pi)^4}\frac{M^2_0F^2(k^2_E)}
{[k^2_E+M^2_0F^2(k^2_E)]^2},
\end{equation}
where $k_E$ stands for the Euclidean four momentum. Taking an input for the constituent-quark mass $M_0$ as $350$ MeV, we determine $\Lambda=1.04$ GeV to reproduce the empirical value for $F_\pi\approx93$ MeV. Note that the value of $\Lambda$ is about $(20\sim30)\%$ larger than the LIM cutoff mass $\sqrt{2}/\bar{\rho}\approx0.8$ GeV, in which $\bar{\rho}\approx1/3$ fm denotes the average instanton size in LIM~\cite{Diakonov:2002fq}. This discrepancy can be understood by that the axial-current conserving terms, which is not taken into account in Eq.~(\ref{eq:FPI}), provides about $30\%$ increase in $F_\pi$~\cite{Nam:2007gf}. Hence, instead of introducing the current-conserving terms to avoid the complexities, we modify the value of $\Lambda$ from the conventional one $\sim0.8$ GeV to reproduce the appropriate empirical values for $F_\mathcal{M}$. It is worth mentioning that the qualitative conclusions of the present work do not change even with the axial-current conserving terms, and those terms will be taken into account in the future works in order to test the quantitative changes. 

Now, we are in a position to take into account the leading-twist PS-meson DA within NLChQM. Since those DAs can be defined in the light-front (LF) formalism for the minimal Fock state for the mesons $\sim{q\bar{q}}$~\cite{Lepage:1980fj}, one must perform a Wick rotation of the effective chiral action via analytical continuation to compute DA~\cite{Praszalowicz:2001wy}. Although the rotation from Euclidean to Minkowski spaces in this model has not been fully proved analytically, we observed that this treatment looks reasonable and can explain many physical quantities qualitatively well such as the moments as shown in our previous works~\cite{Nam:2006au,Nam:2006sx}. Before going further, we define several relations for the LF formalism by defining the light-like vector $n$:
\begin{eqnarray}
\label{eq:LCR}
n_{\mu}&=&(1,0,0,1),\,\,\,\,\bar{n}_{\mu}=(1,0,0,-1),\,\,\,\,n\cdot \bar{n}=2
\cr
k_{\mu}&=&\frac{k^{+}}{2}n_{\mu}+\frac{k^{-}}{2}\bar{n}_{\mu}+k_{T\mu},\,\,\,\,k_{T\mu}=(0,k_1,k_2,0),\,\,\,\,k^{\mu}_{T}=(0,-k_1,-k_2,0)
\cr
k^{+}&=&k_0+k_3=k\cdot n,\,\,\,\,k^{-}=k_0-k_3=k\cdot\bar{n},\,\,\,\,k\cdot k=k^{+}k^{-}-|k_{T}|^2,
\,\,\,\,d^4k=\frac{1}{2}dk^+dk^-dk^2_{T}.
\end{eqnarray}
The PS-meson DA is now defined as follows with the LF formalism for $\mathcal{M}\equiv q\bar{q}'$~\cite{Nam:2006au}:
\begin{equation}
\label{eq:LFWF}
\phi_\mathcal{M}(x)=\frac{1}{i\sqrt{2}F_{q\bar{q}'}}\int^{\infty}_{-\infty}
\frac{d\tau}{2\pi}e^{-i\tau(2x-1)(n\cdot p)}\langle
0 |\bar{\psi}_{q}(\tau n)\rlap{/}{n}\gamma_{5}\exp
\left[ig\int^{\tau}_{-\tau}d\tau'n^{\nu}A_{\nu}(\tau'
n)\right]\psi_{q'}(-\tau n)|(q\bar{q}')(p)\rangle,
\end{equation}
where $p$ denotes the PS-meson four momentum. When we assign $(q,q')=(u,d)$, we have $\pi^+$ as understood. Employing the light-cone gauge $n\cdot A=0$, the exponential term in the r.h.s. of Eq.~(\ref{eq:LFWF}) becomes unity. As shown in detail in our previous works~\cite{Nam:2006au,Nam:2006sx}, DA can be obtained within the nonperturbative (NP) model straightforwardly, resulting in:
\begin{equation}
\label{eq:DA}
\phi^\mathrm{NP}_\mathcal{M}(x)=-\frac{2iN_c}{F^2_{q\bar{q}'}}
\int\frac{d^4k}{(2\pi)^4}\sqrt{M_q(k)M_{q'}(k-p)}\delta\left[\bar{x}p\cdot n-k\cdot n\right]
\frac{[M_{q'}k-M_q(k-p)]\cdot n}{[k^2-M^2_{q}][(k-p)^2-M^2_{q'}]}.
\end{equation}

Here, it is worth mentioning that, in deriving Eq.~(\ref{eq:DA}), we did not consider the momentum dependence in the effective quark mass in the quark propagators to make the problem easy. Moreover, this simplification does not make considerable and qualitative differences in comparison to the full calculations as already shown in Refs.~\cite{Nam:2006au,Nam:2006sx}. Taking into account that DA is obtained by integrating LFWF over $k_\perp$ and using the light-cone identities as shown in Eq.~(\ref{eq:LCR}), we can write the following expression for the PS-meson LFWF for the minimal Fock state from NLChQM:
\begin{eqnarray}
\label{eq:QDAS}
\psi^\mathrm{NP}_\mathcal{M}(x,k^2_\perp)&=&\frac{\bar{x}N_c\Lambda^4\sqrt{M_{q}M_{q'}}\left[xM_{q}+\bar{x}M_{q'}\right]}
{4\pi^3F^2_\mathcal{M}\left[M^2_q-\Lambda^2\right]\left[k^2_\perp+\Lambda^2-x\bar{x}M^2_\mathcal{M}\right]
\left[k^2_\perp+\bar{x}M^2_{q'}+x\Lambda^2-x\bar{x}M^2_\mathcal{M}\right]}
\cr
&+&\frac{\bar{x}N_c\Lambda^4\sqrt{M_{q}M_{q'}}\left[xM_{q}+\bar{x}M_{q'}\right]}
{4\pi^3F^2_\mathcal{M}\left[\Lambda^2-M^2_q\right]
\left[k^2_\perp+xM^2_{q}+\bar{x}\Lambda^2-x\bar{x}M^2_\mathcal{M}\right]
\left[k^2_\perp+xM^2_{q}+\bar{x}M^2_{q'}
-x\bar{x}M^2_\mathcal{M}\right]},
\end{eqnarray}
where $M_\mathcal{M}$ stands for the mass of the PS meson consisting of $\mathcal{M}\sim q\bar{q}'$. We also introduced a notation $\bar{x}=1-x$. The normalization condition for DA in Eq.~(\ref{eq:DAVDA}) will be used to determine the model parameters $\Lambda$ with $F_\mathcal{M}$ as an input. If we take the chiral limit (CL), i.e. $M_{q}=M_{q'}=M_0$ and $M_{q\bar{q}'}=M_\pi=0$, the pion LFWF is simplified further, resulting in:
\begin{eqnarray}
\label{eq:QDACL}
\psi^\mathrm{NP,CL}_\pi(x,k^2_\perp)
&=&\frac{x\bar{x}N_c\Lambda^4M^2_0\left(k^2_\perp+\frac{M^2_0+\Lambda^2}{2}\right)}
{2\pi^3F^2_\pi\left[k^2_\perp+\Lambda^2\right]\left[k^2_\perp+M^2_0\right]
\left[k^2_\perp+x\Lambda^2+\bar{x}M^2_0\right]
\left[k^2_\perp+\bar{x}\Lambda^2+xM^2_0\right]},
\end{eqnarray}
which is the same with Eq.~(18) in Ref.~\cite{Petrov:1998kg} for $p_\perp\approx0$. From Eq.~(\ref{eq:QDACL}), it is straightforward to obtain the analytic expression for the pion DA in the chiral limit by integrating LFWF over $k_\perp$:
\begin{equation}
\label{eq:DACL}
\phi^\mathrm{NP,CL}_\pi(x)=\int^\infty_0dk^2_\perp\,\psi^\mathrm{NP,CL}_\pi(x,k^2_\perp)
=\frac{N_c\Lambda^4M^2_0}{4\pi^2F^2_\pi(\Lambda^2-M^2_0)^2}
\ln\left[\frac{(x\Lambda^2+\bar{x}M^2_0)(\bar{x}\Lambda^2+xM^2_0)}{\Lambda^2M^2_0}\right],
\end{equation}
where $\Lambda>M_0$. As mentioned previously, we have assumed that $\Psi^\mathrm{NP}_\mathcal{M}(x,k^2_\perp)=\psi^\mathrm{NP}_\mathcal{M}(x,k^2_\perp)$ as in Eq.~(\ref{eq:MOMENT5}). If this is the case, QDA can be derived analytically from Eq.~(\ref{eq:QDATMDA}) by integrating TMDA over $x$ and $k_1$. In order to investigate the current-quark mass effects in QDA,  we expand QDA with respect to the mass difference between the quarks inside the PS meson, i.e. $\Delta M = |M_q-M_{q'}|=|m_q-m_{q'}|\ll\Lambda$. Finally, we arrive at an  expression for QDA from NLChQM up to $\mathcal{O}\left(\Delta M\right)$ as follows:
\begin{eqnarray}
\label{eq:QDASIM}
Q^\mathrm{NP}_\mathcal{M}(y,p_3)
&=&\frac{N_cM^2_0\Lambda^4}{8\pi^2 F^2_\pi\eta^4}\Bigg\{
\ln\left[\frac{
\left[\eta^2+2p_3f_-(\bar{y},\Lambda)\right]^2
\left[\eta^2-2p_3f_-(\bar{y},M_0)\right]^2f_+(\bar{y},\Lambda)\,
f_+(\bar{y},M_0)\,
f_+(y,\Lambda)\,
f_+(y,M_0)}
{\left[\eta^2+2p_3f_+(y,\Lambda)\right]^2
\left[\eta^2-2p_3f_+(y,M_0)\right]^2f_-(\bar{y},\Lambda)\,
f_-(\bar{y},M_0)\,
f_-(y,\Lambda)\,
f_-(y,M_0)}\right]
\cr
&-&\frac{\Delta M}{M_0}
\Bigg[(2y-3)
\ln\left[
\frac{
f_-(\bar{y},\Lambda)f_-(\bar{y},M_0)
[\eta^2+2p_3f_+(y,\Lambda)]
[\eta^2-2p_3f_+(y,M_0)]}
{f_+(y,\Lambda)f_+(y,M_0)
[\eta^2+2p_3f_-(\bar{y},\Lambda)]
[\eta^2-2p_3f_-(\bar{y},M_0)]}
\right]
\cr
&+&
\frac{\eta^2}{p^2_3}\ln\left[
\frac{[\eta^2+2p_3f_+(y,\Lambda)][\eta^2+2p_3f_-(\bar{y},\Lambda)]}
{[\eta^2-2p_3f_+(y,M_0)][\eta^2-2p_3f_-(\bar{y},M_0)]}\right]
\Bigg]\Bigg\}+\mathcal{O}\left(\Delta M^2\right)
\end{eqnarray}
where we defined a function as follows for brevity:
\begin{equation}
\label{eq:FUNC}
f_\pm(y,M_0)=xp_3\pm\sqrt{M_0^2+y^2p^2_3},
\end{equation}
with $\bar{y}=1-y$ and $\eta^2=\Lambda^2-M^2_0>0$. We have $\Delta M=(0,135)$ MeV for the pion and kaon, respectively. Note that this expression for QDA depends on the regularization scheme and the model scale $\Lambda\sim1$ GeV in the present work. Since we are interested only in the low-energy region, we will drop the superscript NP hereafter.

\section{Numerical results and discussions}
In this Section, we provide numerical results for the various amplitudes and relevant discussions. Before showing the numerical results, we list all the relevant parameters in Table~\ref{TABLE1}. There are three input values, i.e. the constituent-quark mass $M_0$ and current-quark mass for the light and strange quarks $(m_{u,d},m_s)$ in the isospin symmetry. Using the empirical and experimental values for $M_{\pi,K}$ and $F_{\pi,K}$ as shown there, we determine the model scale parameter $\Lambda$ to satisfy the normalization condition for DAs in Eq.~(\ref{eq:DAVDA}). For the cases of the pion in the chiral limit (CL), the pion and kaon with the finite current-quark masses, the scale parameter becomes about $1$ GeV as observed already in Ref.~\cite{Nam:2006au}.

\begin{table}[b]
\begin{tabular}{c||c|c|c|c|c}
&$M_0=350$ MeV&$m_{u,d}=5$ MeV&$m_s=135$ MeV&\\
\hline
\hline
Pion at CL&$F_\pi=93$ MeV&$M_q=M_0$&$M_{q'}=M_0$&$M_\pi=0$ MeV&$\Lambda=1.02$ GeV\\
\hline
Pion&$F_\pi=93$ MeV&$M_q=(m_{u,d}+M_0)=355$ MeV&$M_{q'}=(m_{u,d}+M_0)=355$ MeV&$M_\pi=140$ MeV&$\Lambda=1.01$ GeV\\
\hline
Kaon&$F_K=113$ MeV&$M_q=(m_{u,d}+M_0)=355$ MeV&$M_{q'}=(m_s+M_0)=485$ MeV&$M_K=495$ MeV&$\Lambda=1.05$ GeV\\
\end{tabular}
\caption{Model parameters for the present calculations. With these values, the pion and kaon DAs satisfy the normalization condition, i.e. $\int dx\,\phi_{\pi,K}(x)=1$.}
\label{TABLE1}
\end{table}

\begin{figure}[t]
\includegraphics[width=8.5cm]{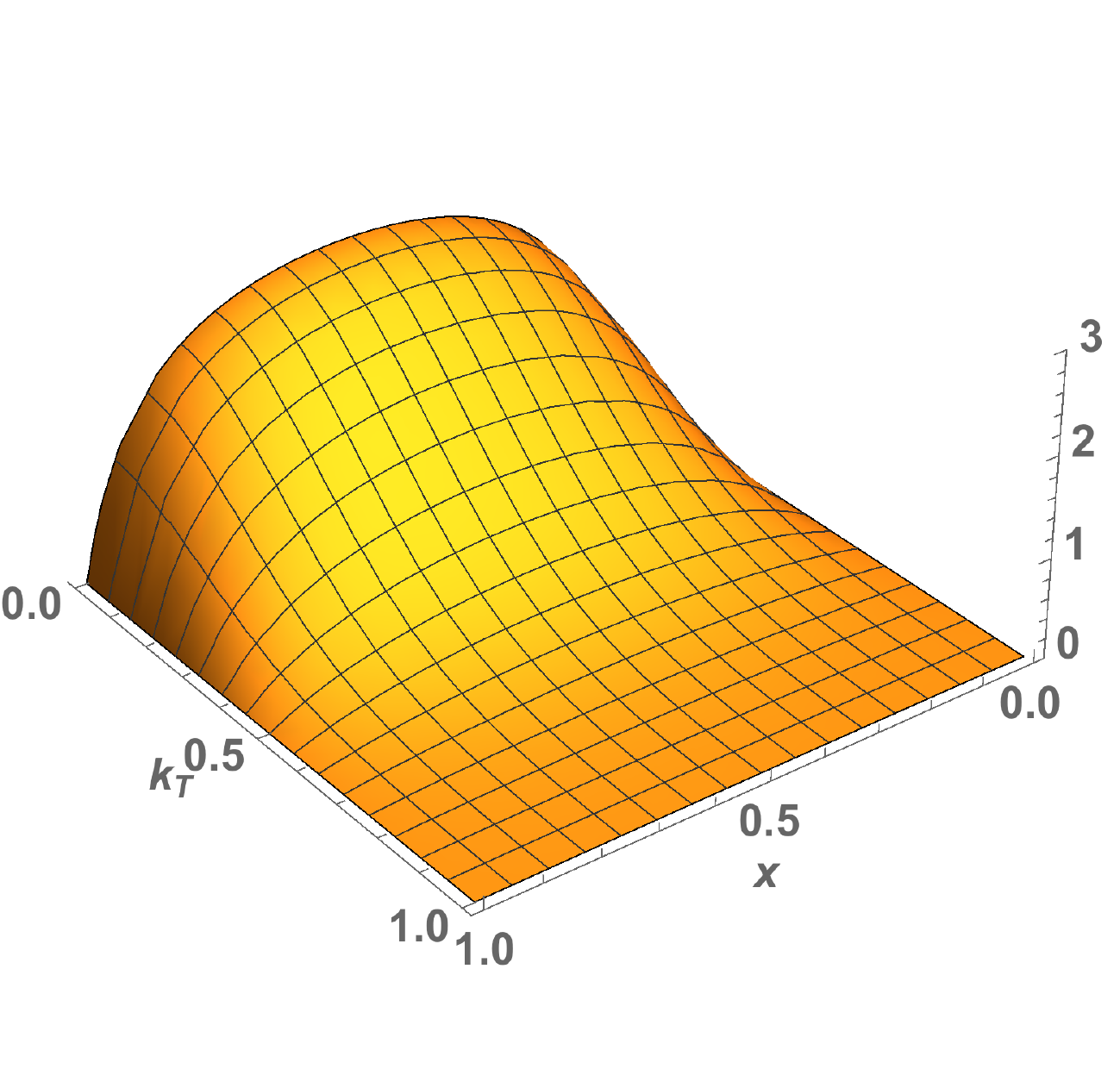}
\includegraphics[width=8.5cm]{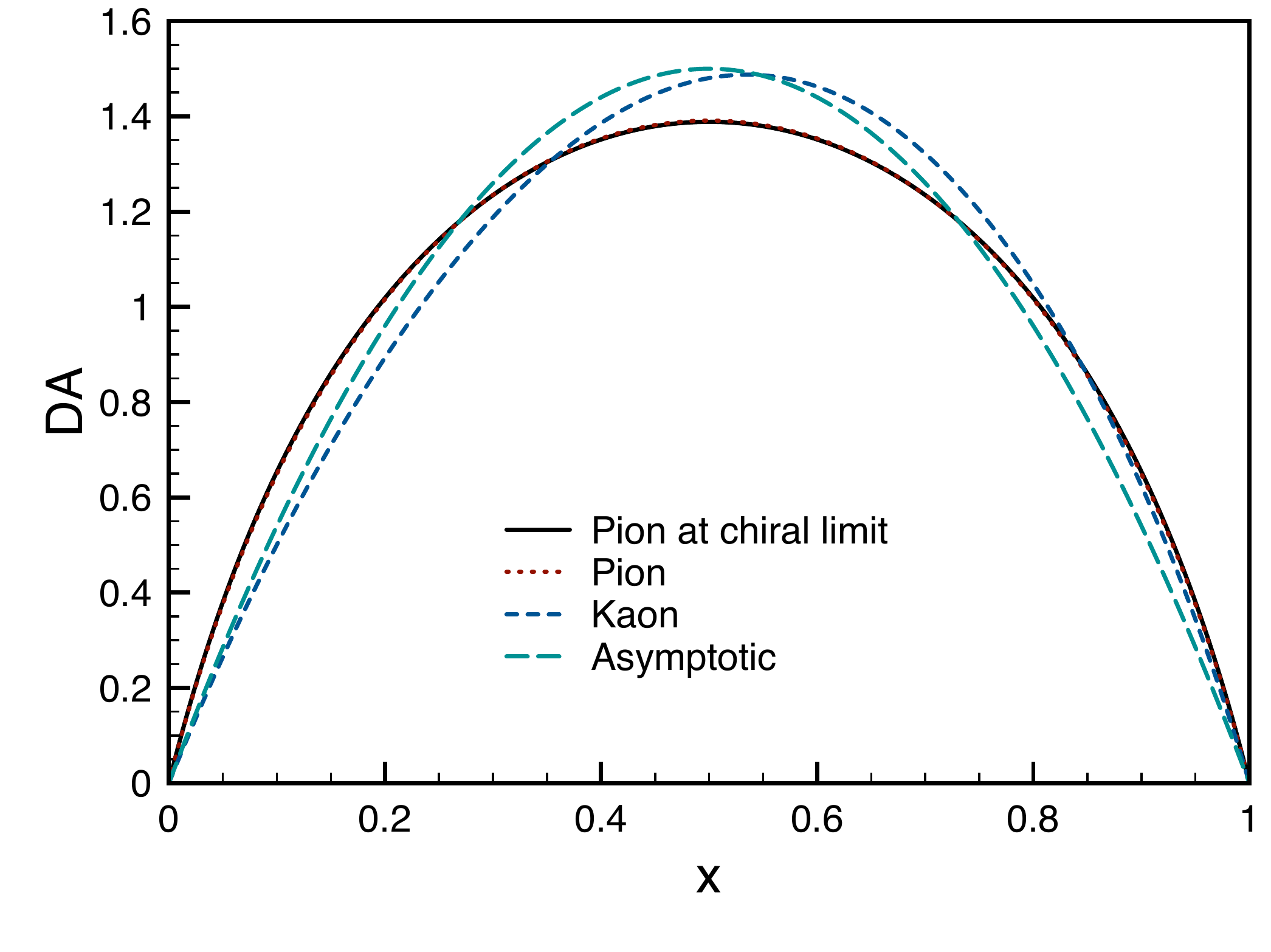}
\caption{(Color online) Left: TMDA (or LFWF) for the pion in the chiral limit $\psi^\mathrm{CL}_\pi(x,k^2_\perp)$ in Eq.~(\ref{eq:QDACL}) as a function of $x$ and $k_\perp\equiv k_T$ [GeV] from NLChQM. We will take this as TMDA $\Psi^\mathrm{CL}_\pi(x,k^2_\perp)$ at the nonperturbative region as discussed in the text. Right: DAs in Eq.~(\ref{eq:DA}) as functions of $x$ from NLChQM for the pion at chiral limit (solid), pion (dotted), and kaon (dashed) cases. Asymptotic DA is also given for comparison (long-dashed). See the text for details.}
\label{FIGTMDADA}
\end{figure}

Using those parameters, in the left panel of Fig.~\ref{FIGTMDADA}, we first show the numerical result for TMDA (or LFWF) as a function of $x$ and $k_\perp$ for the pion in the chiral limit. As shown there, it is symmetric with respect to $x$ due to the fact that $\Delta M=0$. We also observe that the curve decreases as a function of $k_\perp$ and becomes almost zero beyond $k_\perp\approx0.7$ GeV, indicating the nonperturbative nature of TMDA $\Psi^\mathrm{CL}_\pi(x,k^2_\perp)$ from NLChQM. Note that, in Ref.~\cite{Radyushkin:2017gjd}, several model parameterizations for TMDA were given: The Gaussian and Bessel function types. We find that our TMDA based on LFWF of NLChQM is rather similar to the Bessel one. In the right panel of Fig.~\ref{FIGTMDADA}, we depict the meson DAs for the pion at the chiral limit $\phi^\mathrm{CL}_\pi$ (solid), pion $\phi_\pi(x)$ (dotted) and kaon $\phi_K(x)$ (dashed) with the finite current-quark masses, by integrating LFWF in Eq.~(\ref{eq:QDAS}) over $k_\perp$. As shown there, the pion DAs are almost the same and symmetric with respect to $x$ as expected: $m_u=m_d\sim0$~\cite{Nam:2006au}. Due to the considerable difference in the strange- and light-quark masses $\Delta M\ne0$, the kaon DA becomes asymmetric. For comparison, we also show the asymptotic DA $\phi_\mathrm{asym}(x)=6x(1-x)$ in the same panel. It is obvious that our nonperturbative results for DAs are different from the asymptotic one. Note that all these observations for DAs are consistent with our previous results~\cite{Nam:2006au}.

\begin{figure}[t]
\includegraphics[width=8.5cm]{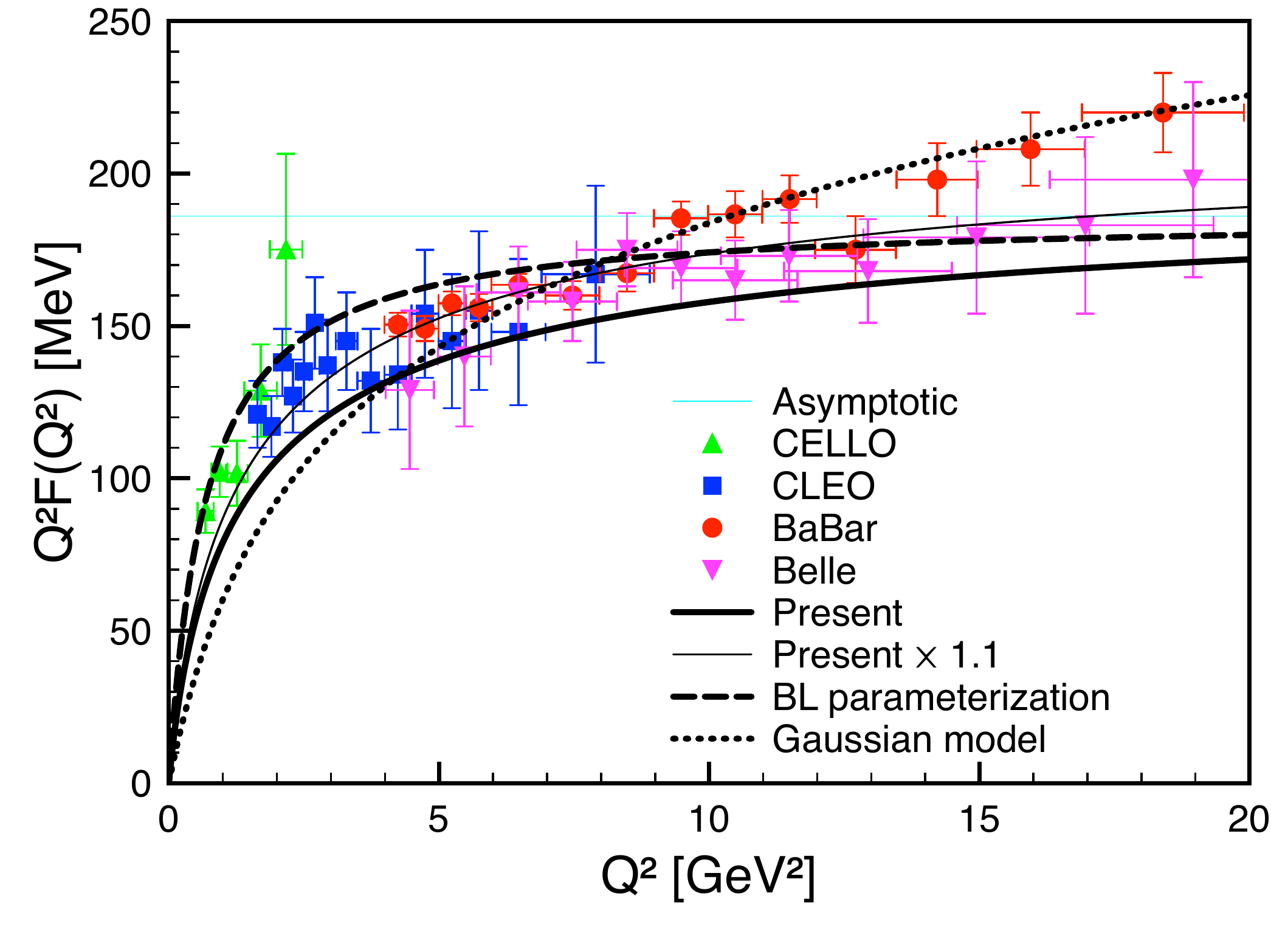}
\caption{(Color online) $\gamma^*\gamma\to\pi^0$ transition form factor derived from TMDA as a function of $Q^2$. The solid, dashed, and dotted lines denote the present calculation, the Brodsky-Lepage (BL) parameterization~\cite{Oganesian:2015ucv}, and the Gaussian model~\cite{Radyushkin:2015gpa}, respectively, whereas the horizontal line denotes the asymptotic value $2F_\pi\approx186$ MeV. The experimental data are taken from CELLO~\cite{Behrend:1990sr}, CLEO~\cite{Gronberg:1997fj}, BaBar~\cite{Aubert:2009mc}, and Belle~\cite{Uehara:2012ag}.}
\label{FIGFF}
\end{figure}

In Ref.~\cite{Lepage:1980fj,Radyushkin:2015gpa}, it was given that the pion-photon transition form factor (FF) is related to TMDA as follows:
\begin{equation}
\label{eq:FF}
F_{\gamma\gamma^*\pi^0}(Q^2)=\frac{2F_\pi}{3}\int^1_0\frac{dx}{xQ^2}\int^{xQ^2}_0
\frac{dk^2_\perp}{xQ^2}\int^{k_\perp}_0d^2p_\perp\,\Psi(x,p^2_\perp).
\end{equation}
In the asymptotic limit, Eq.~(\ref{eq:FF}) is approximated to
\begin{equation}
\label{eq:FF1}
F^\mathrm{asym}_{\gamma\gamma^*\pi^0}(Q^2)\approx\frac{2F_\pi}{3}\int^1_0\frac{dx}{xQ^2}\phi_\mathrm{asym}(x),
\end{equation}
and it turns out that $Q^2F^\mathrm{asym}_{\gamma\gamma^*\pi^0}(Q^2)=2F_\pi\approx0.186$ GeV for $F_\pi\approx93$ MeV as understood by the Brodsky-Lepage limit~\cite{Lepage:1980fj}. Within our model, TMDA is equivalent to LFWF at the low-renormalization scale as mentioned previously, we can compute the transition FF in the chiral limit, using Eqs.~(\ref{eq:QDACL}) and (\ref{eq:FF}), and the following integral, which appears in the r.h.s. of Eq.~(\ref{eq:FF}):
\begin{equation}
\label{eq:}
\int^{k_\perp}_0d^2p_\perp\Psi^\mathrm{CL}_\pi(x,p^2_\perp)
=\frac{N_c\Lambda^4M^2_0}{4\pi^2F^2_\pi(\Lambda^2-M^2_0)^2}
\ln\left[\frac{(\Lambda^2+k^2_\perp)(M^2_0+k^2_\perp)(x\Lambda^2+\bar{x}M^2_0)(\bar{x}\Lambda^2+xM^2_0)}{\Lambda^2M^2_0(x\Lambda^2+\bar{x}M^2_0+k^2_\perp)(\bar{x}\Lambda^2+xM^2_0+k^2_\perp)}\right].
\end{equation}
By doing that, the numerical result of NLChQM for the transition FF multiplied by $Q^2$ is given in Fig.~\ref{FIGFF}  as a function of $Q^2$ in the thick-solid line. The thick-dashed and thick-dotted lines indicate numerical results from the Brodsky and Lepage (BL) parameterization~\cite{Oganesian:2015ucv} and the Gaussian-model (GM) TMDA with the flat DA, i.e. $\phi_\pi(x)=1$~\cite{Radyushkin:2015gpa}:
\begin{equation}
\label{eq:LBPARA}
F^\mathrm{BL}_{\gamma\gamma^*\pi^0}(Q^2)=\frac{2F_\pi Q^2}{Q^2+8\pi^2F^2_\pi},\,\,\,\,
F^\mathrm{GM}_{\gamma\gamma^*\pi^0}(Q^2)=
\frac{2F_\pi}{3}\int^1_0\,\frac{dx}{xQ^2}\left[1-\frac{\Lambda'^2}{xQ^2}
\left(1-e^{-\frac{xQ^2}{\Lambda'^2}}\right)\right],
\end{equation}
where $\Lambda'^2=0.35\,\mathrm{GeV}^2$. The thin horizontal line denotes the asymptotic value $\sim186$ MeV. We also show the experimental data from CELLO~\cite{Behrend:1990sr}, CLEO~\cite{Gronberg:1997fj}, BaBar~\cite{Aubert:2009mc}, and Belle~\cite{Uehara:2012ag} for comparison.

As shown there, although the theoretical curve seems to undershoot the data slightly within the error bars, the qualitative behavior and strength of the curve coincides with the data relatively well. Naturally, our nonperturbative result can not reproduce the asymptotic value at the high-$Q^2$ region. Since our parameters are determined self-consistently within the model, we have almost no rooms to tune the parameters. However, if we admit about $10\%$ allowance in the numerical results, we can obtain a seemingly better curve as shown in the thin-solid line in the figure. By comparing the present result and that of GM with the flat DA for the pion, in the nonperturbative region, non-flat DA must be favorable, while DA approaches to the flat one as $Q^2$ increases in order to describe the data for the whole range $Q^2=(0\sim20)\,\mathrm{GeV}^2$, including the BaBar data.

In the zero-virtuality limit $Q^2\to0$, we have the following value numerically:
\begin{equation}
\label{eq:ZVL}
\lim_{Q^2\to0}F^\mathrm{NLChQM}_{\gamma\gamma^*\pi^0}(Q^2)=0.191\,\mathrm{GeV}^{-1}.
\end{equation}
Note that the Adler-Bell-Jackiw axial anomaly gives nonzero values of the transition form factors at $Q^2=0$ for the case with two real photons: $F_{\gamma\gamma\pi^0}(0)=(4\pi^2F_\pi)^{-1}\approx0.272\,\mathrm{GeV}^{-1}$~\cite{RuizArriola:2006jge,Oganesian:2015ucv}. From this and Eq.~(\ref{eq:ZVL}), the ratio becomes
\begin{equation}
\label{eq:RATIO}
F^\mathrm{NLChQM}_{\gamma\gamma^*\pi^0}(0)/F_{\gamma\gamma\pi^0}(0)=0.702.
\end{equation}
In Ref.~\cite{Lepage:1980fj}, the ratio was discussed, resulting in exactly $0.5$ by considering the minimal $q\bar{q}$ Fock state in the LF frame for the meson. It also turned out that a model TMDA with the Gaussian distribution of $k_\perp$ gives the ratio $0.53$ with $\Lambda'^2=0.2\,\mathrm{GeV}^2$~\cite{Radyushkin:2015gpa}.

\begin{figure}[t]
\includegraphics[width=8.5cm]{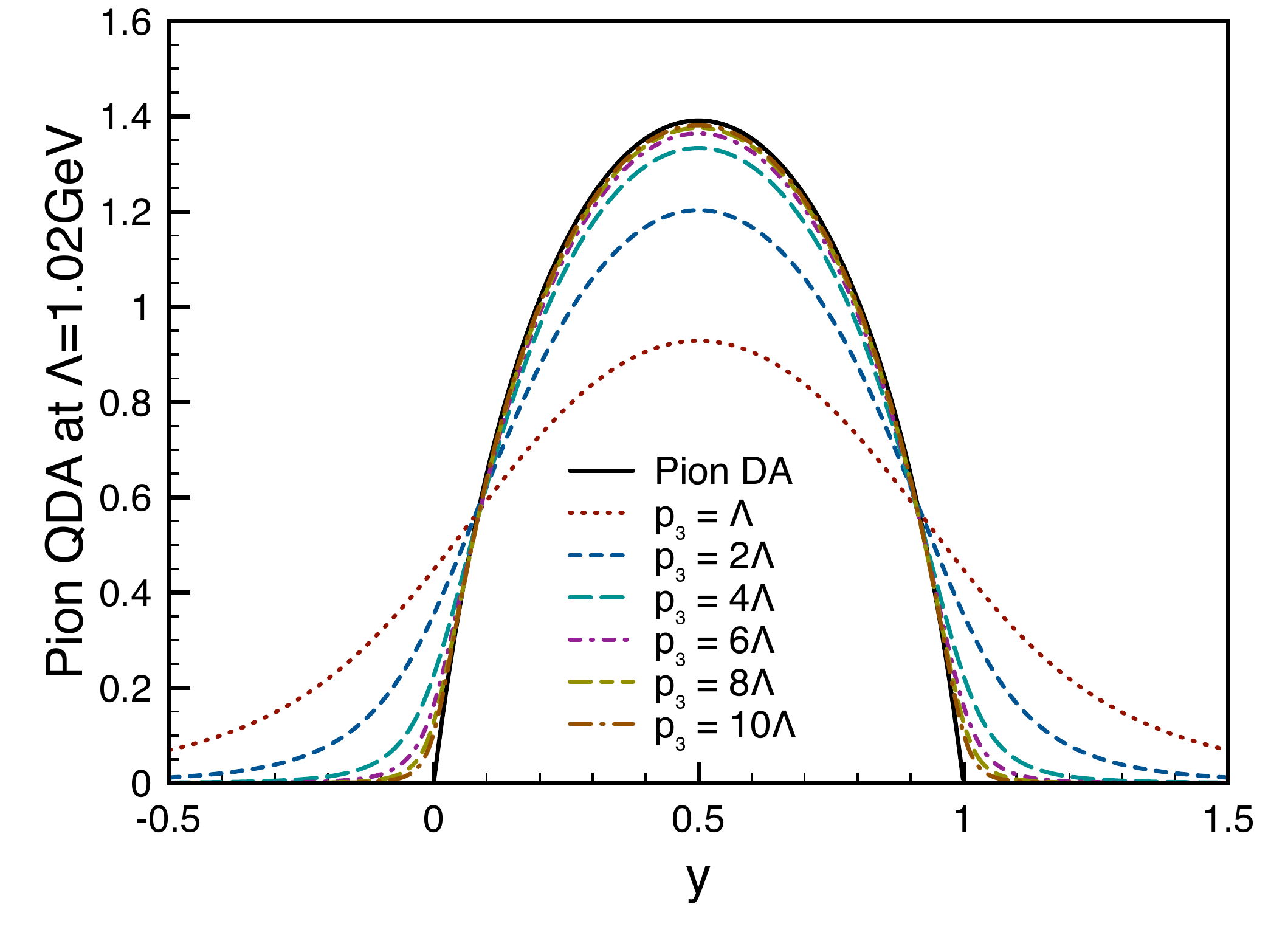}
\includegraphics[width=8.5cm]{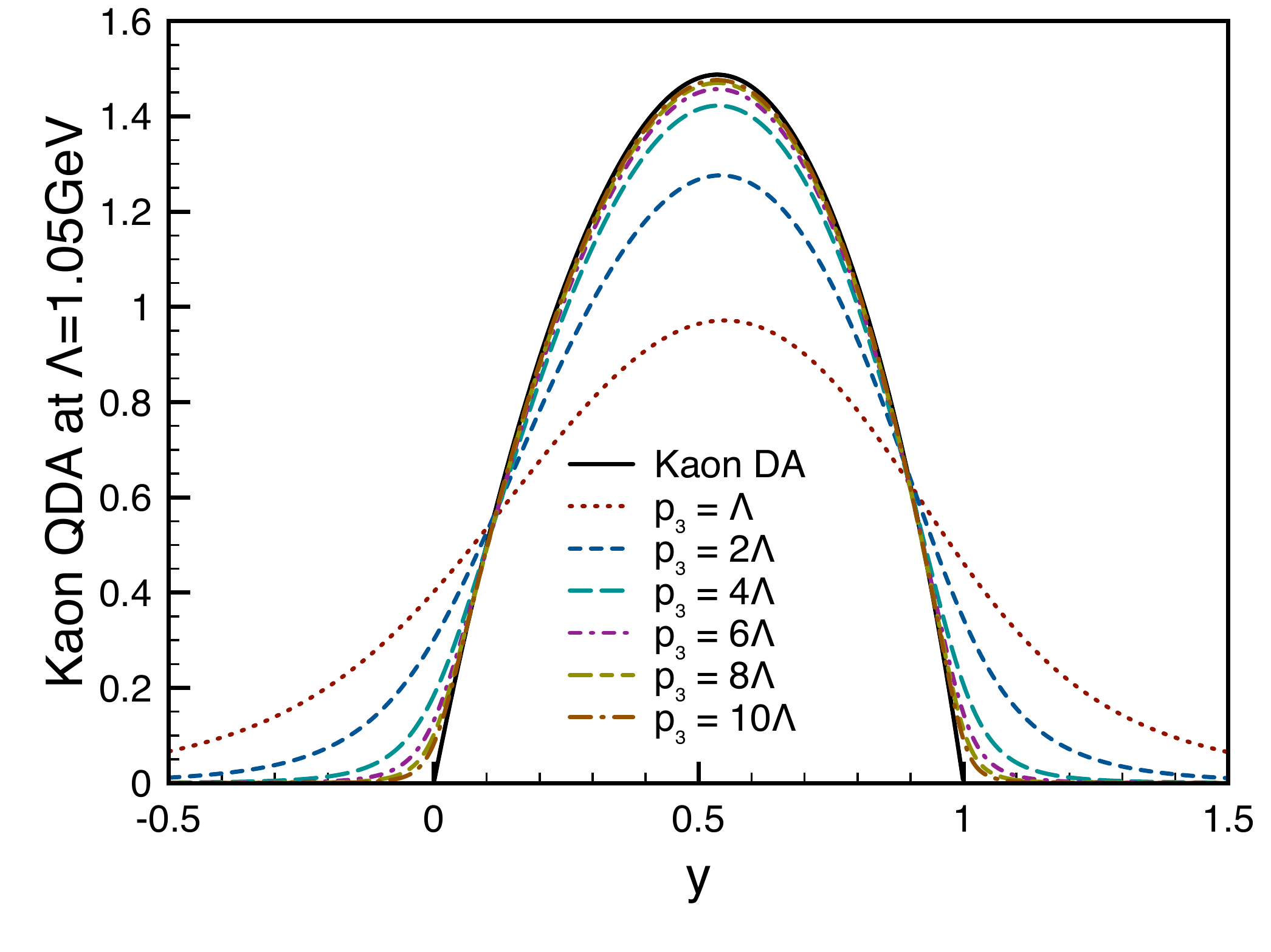}
\caption{(Color online) Pion (left) and kaon (right) QDAs as functions of $y$ for various $p_3$ with $\Lambda=1.02$ GeV and $1.05$, GeV, respectively. We also show pion and kaon DAs for comparison.}
\label{FIGQDA}
\includegraphics[width=8.5cm]{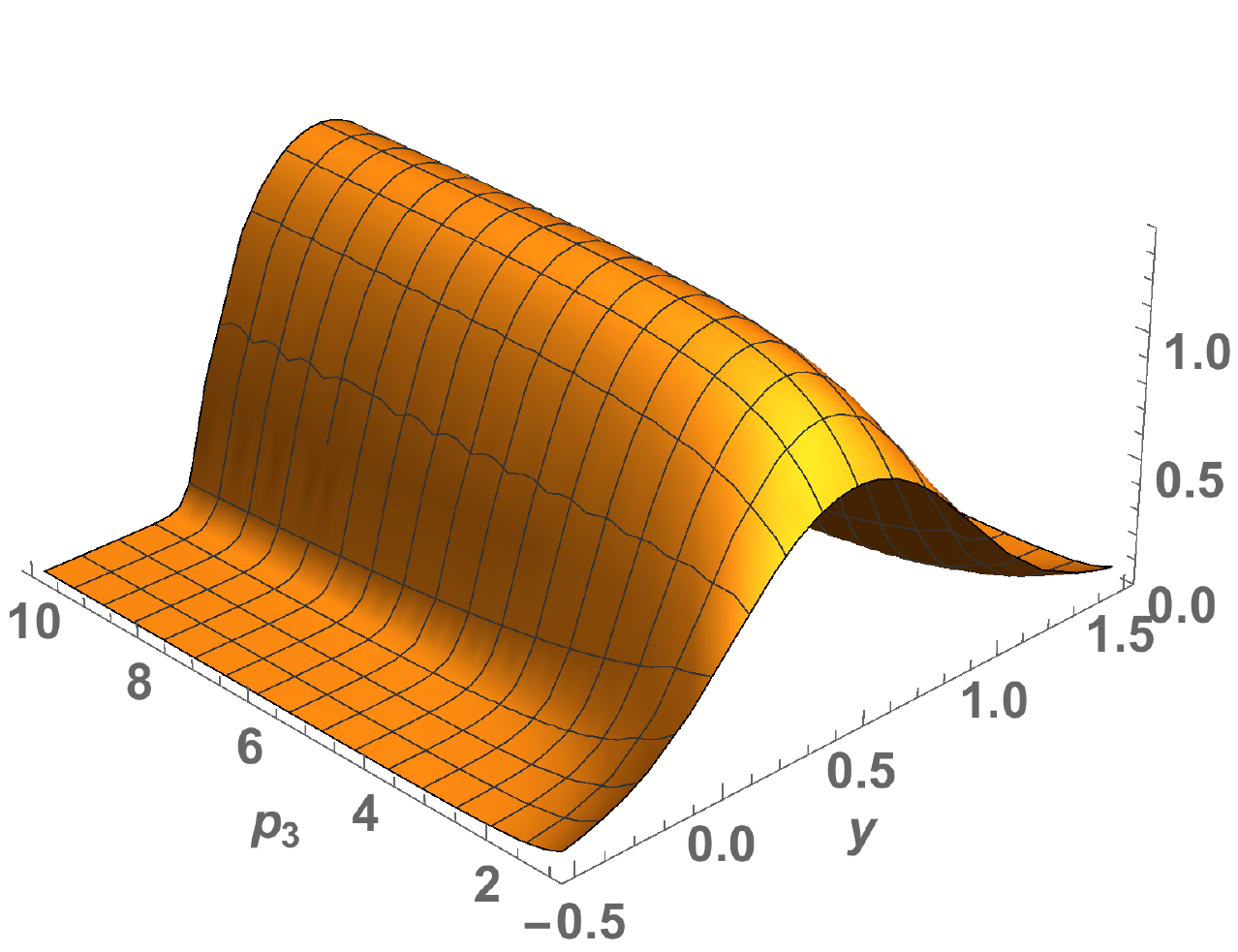}
\caption{(Color online) QDA in the chiral limit in Eq.~(\ref{eq:QDASIM}) as a function of $y$ and $p_3$ [GeV].}
\label{FIGQDA3D}
\end{figure}

After confirming our numerical results for TMDA and DA, we are now in a position to explore QDA in the limit of $p_3\to\infty$. Using Eq.~(\ref{eq:QDASIM}), we draw QDAs as functions of $y$ for different $p_3$ values for the pion (left) and kaon (right) in Fig.~\ref{FIGQDA}. For simplicity, we choose $p_3=n\Lambda$ for $n=(1\sim10)$. The values for $\Lambda$ for the pion and kaon are $1.02$ GeV and $1.05$ GeV, respectively. For comparison, We also depict the pion and kaon DA, computed directly from LFWF. It is shown that the curve of QDA approaches to that of DA as $p_3$ increases, as expected. Even at $p_3\approx10$ GeV, the QDA curve does not match well with DA, being apparent at the end points $x=(0,1)$, although qualitative behavior of the curves look similar to each other. As for the kaon, we observe that the asymmetric DA curve is well reproduced from QDA. It is interesting that the inflection points of the QDA curves at $y\approx(0.1,0.9)$ seem to remain the same for different $p_3$ values for both the pion and kaon. In Fig.~\ref{FIGQDA3D}, we show the numerical results for QDA as a function of $p_3$ and $y$ for the pion. It is clearly shown that, as $p_3$ increases, the end-point values approach to zero. We verified that the difference between QDA and DA disappears qualitatively  for $p_3\gtrsim30$ GeV.

\begin{figure}[t]
\includegraphics[width=8.5cm]{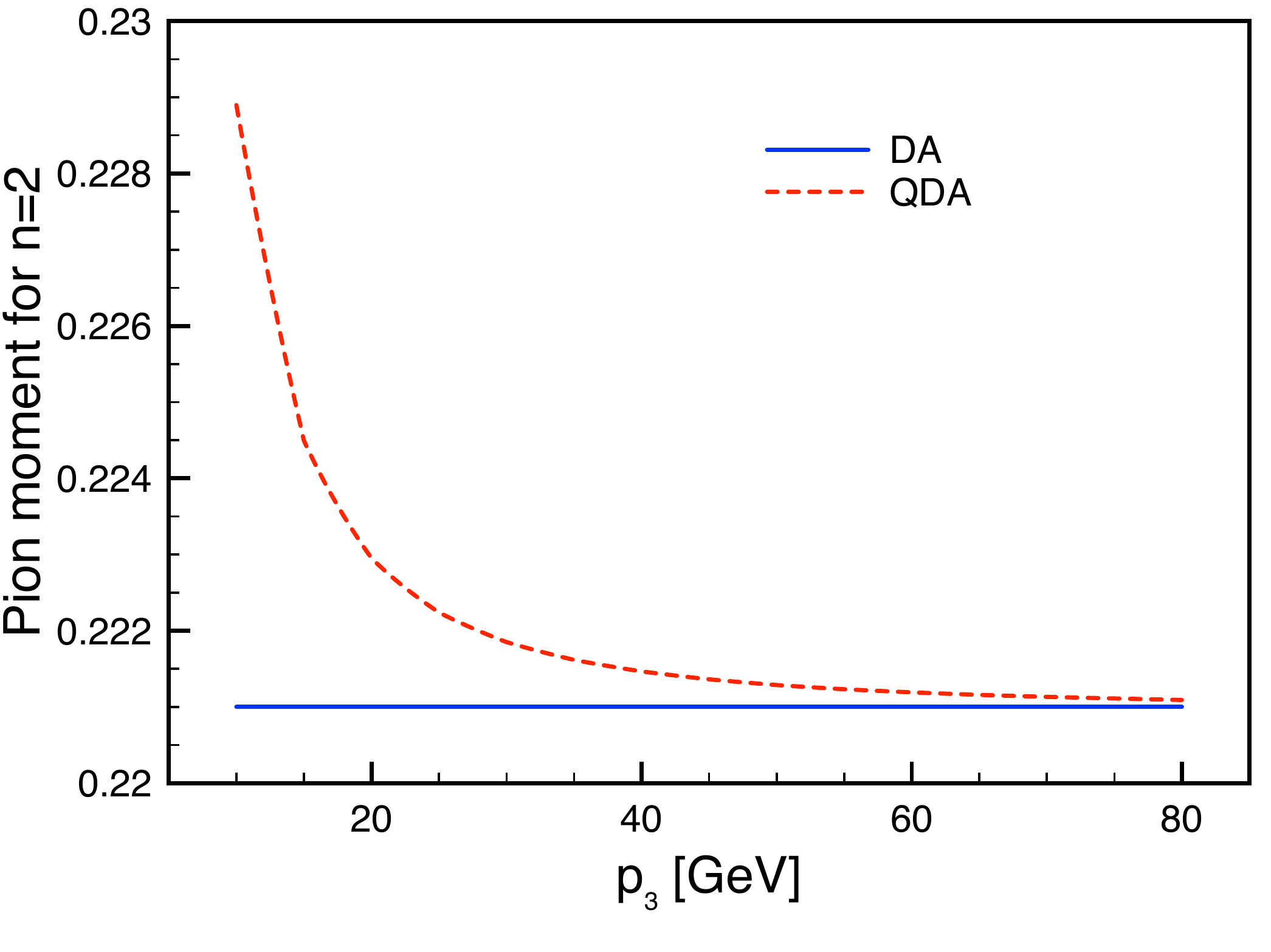}
\includegraphics[width=8.5cm]{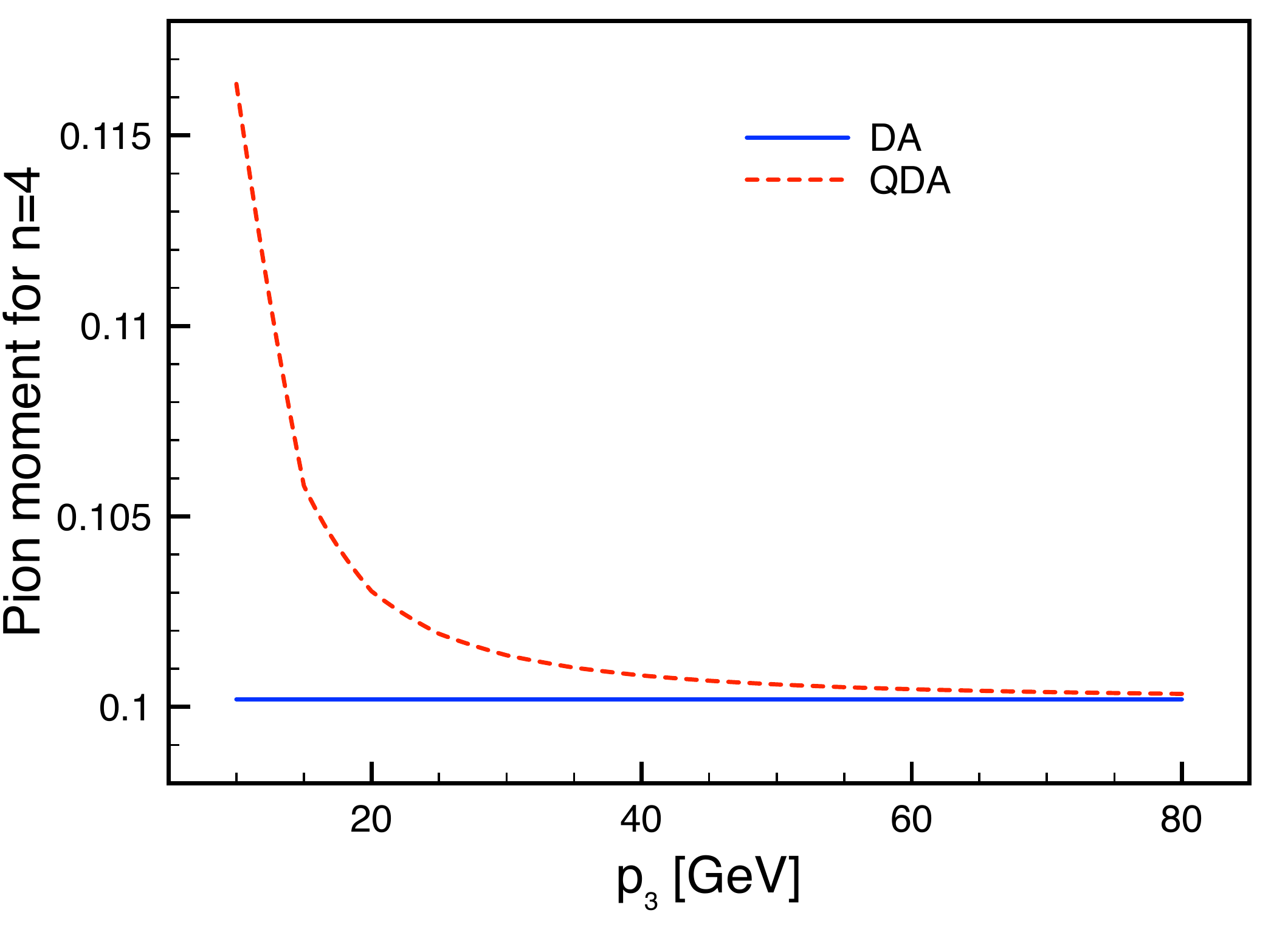}
\caption{(Color online) Pion moments $\langle\xi^n\rangle_\pi$ for $n=2$ (left) and $n=4$ (right) obtained from pion QDA (dashed) as a function of $p_3$ and DA. The horizontal solid line indicates that directly computed from DA.}
\label{FIGMOM}
\end{figure}

Finally, in order to confirm the equivalence between DA and QDA at $p_3\to\infty$ in the present model, we want to compute the pion- and kaon-DA moments $\langle\xi^n\rangle$ for $\xi\equiv(y-\bar{y})=2x-1$ from DA and QDA as follows:
\begin{equation}
\label{eq:MOM}
\langle\xi^n\rangle^\mathrm{DA}_\mathcal{M}=\int^1_0 (2x-1)^n\phi_\mathcal{M}(x)dx,\,\,\,\,
\langle\xi^n\rangle^\mathrm{QDA}_\mathcal{M}=\lim_{p_3\to\infty}\int^\infty_{-\infty}(2y-1)^nQ_\mathcal{M}(y,p_3)dy.
\end{equation}
In Fig.~\ref{FIGMOM}, we draw the moments computed from QDA as a function of $p_3$ for $n=2$ (left) and $n=4$ (right) for the pion in the dashed line. For comparison, we also show the moments directly obtained from DA in the horizontal solid line. Up to $p_3\approx30$, the curves of the QDA moments decreases stiffly with respect to $p_3$, then smoothly gets close to that from DA. Beyond $p_3\approx80$ GeV, there appears only negligible difference between the moments from QDA and DA as expected. In Table~\ref{TABLE2}, we list the numerical values of the moments for various cases: Those from DA and from QDA at different $p_3$ values for the pion and kaon. It turns out that the $p_3$ dependence of the QDA moments becomes much more crucial for the higher moments. For instance, the kaon QDA moment for $n=1$ does not depend on the values of $p_3$, $\langle\xi\rangle^\mathrm{QDA}_K$=0.0277, whereas there is about $10\%$ difference for $p_3=(10\sim30)\Lambda$, $\langle\xi^4\rangle^\mathrm{QDA}_K=(0.0898\sim0.1034)$. A similar tendency is also observed for the pion. Nonetheless, for $p_3\approx30$ GeV, the discrepancy between the DA and QDA moments becomes only a few percent, i.e. two moments are the same from a practical point of view. In other words, from a nonperturbative approach, one should take $p_3\gtrsim30\Lambda$ more or less to obtain practically reasonable results for the moments from QDA, where $\Lambda$ denotes a nonperturbative scale of the approach, although it depends on the regularization scheme.

\begin{table}[b]
\begin{tabular}{c|cccc||c|cccc}
&$n=1$&$n=2$&$n=3$&$n=4$&&$n=1$&$n=2$&$n=3$&$n=4$\\
\hline
$\langle\xi^n\rangle^\mathrm{DA}_\pi$&$-$&$0.2210$&$-$&$0.1002$
&$\langle\xi^n\rangle^\mathrm{DA}_K$&$0.0277$&$0.2043$&$0.0122$&$0.0887$\\
\hline
$\langle\xi^n\rangle^\mathrm{QDA}_\pi$ at $p_3=10\Lambda$&$-$&$0.2287$&$-$&$0.1159$
&$\langle\xi^n\rangle^\mathrm{QDA}_K$ at $p_3=10\Lambda$&$0.0277$&$0.2118$&$0.0120$&$0.1034$\\
\hline
$\langle\xi^n\rangle^\mathrm{QDA}_\pi$ at $p_3=20\Lambda$&$-$&$0.2229$&$-$&$0.1030$
&$\langle\xi^n\rangle^\mathrm{QDA}_K$ at $p_3=20\Lambda$&$0.0277$&$0.2062$&$0.0121$&$0.0913$\\
\hline
$\langle\xi^n\rangle^\mathrm{QDA}_\pi$ at $p_3=30\Lambda$&$-$&$0.2218$&$-$&$0.1013$
&$\langle\xi^n\rangle^\mathrm{QDA}_K$ at $p_3=30\Lambda$&$0.0277$&$0.2052$&$0.0122$&$0.0898$\\
\end{tabular}
\caption{The moments from the pion and kaon DA and QDA for $\xi=2x-1$ and $2y-1$, respectively.}
\label{TABLE2}
\end{table}
\section{Summary and conclusion}
In the present work, we investigated the PS-meson quasi-distribution amplitudes (QDA) employing the nonlocal chiral-quark model (NLChQM). We first computed the transverse-momentum distribution amplitude (TMDA) for the pion and kaon, using the knowledge of the light-front wave function (LFWF) within the model, then converted them into QDA by replacing the transverse momentum $k^2_\perp\to k^2_1+(x-y)^2p^2_3$, where $p_3$ stands for the PS-meson momentum in the longitudinal direction and by integrating it over $k_1$ and $x$. All the model parameters were determined using the DA normalization condition and to reproduce the PS-meson weak-decay constants. Below, we summarize the relevant observations in the present work:
\begin{itemize}
\item Assuming a specific choice of the integration kernel, i.e. a delta function, we derive a relation between TMDA and LFWF at nonperturbative region, i.e. $\Psi^\mathrm{NP}_\mathcal{M}(x,k^2_\perp)=\psi^\mathrm{NP}_\mathcal{M}(x,k^2_\perp)$ as a trial, although this choice is not unique but the simplest. By doing that, QDA is extracted from TMDA.
\item LFWF at the nonperturbative scale $\Lambda\approx1$ GeV is obtained from NLChQM, which is based on the liquid-instanton QCD-vacuum model, and all the model parameters are determined by the relevant conditions, such as the normalization of the PS-meson DAs.
\item Once we have QDA from NLChQM in hand, an analytic expression for QDA for the pion and kaon are obtained in terms of the current-quark mass difference $\Delta M=|M_q-M_{\bar{q}}|\propto\Delta m_q$ at the nonperturbative region with $\Lambda\approx1$ GeV.
\item The transition form factor is also computed using the model TMDA, resulting in qualitative agreement with the data. This observation supports our consideration that TMDA is equivalent to LFWF at the low-energy scale. 
\item We verified numerically that QDA approaches to DA as $p_3$ increases as expected. Note that the symmetric and asymmetric shapes for the pion and kaon DAs are well reproduced by QDA at $p_3\to\infty$. More or less, $p_3\gtrsim80$ GeV, the difference between the QDA and DA becomes almost negligible, they become the same practically.
\item The moments $\langle\xi^n\rangle_{\pi,K}$ are also computed using DA as well as QDA. It turns out that the higher QDA moments are more sensitive to the change of $p_3$, whereas the lower ones depend less on it. At $p_3=30\Lambda\approx30$ GeV, the moments show only a few percent differences, when those from DA and QDA are compared.
\end{itemize}

Although our consideration, i.e., TMDA $\sim$ LFWF at the nonperturbative region, is not fully proved for the wider energy regions, it is a good starting point to investigate QDA and its interesting features as a first trial, considering the analogous behaviors between them. As reported in Refs.~\cite{Nam:2006au,Nam:2006sx}, the end-point behaviors are deeply affected by the form-factor types. For instance, if we change the power of the form factor in Eq.~(\ref{eq:EFFM}) from {\it two} to {\it three}, the end-point curves of DA becomes more curvy. Thus, it seems interesting for us to see the effects of different form-factor schemes. Moreover, in Ref.~\cite{Nam:2006sx}, the axial-vector current conservation was explored for computing DAs and will give additional terms to the present QDA. As suggested in Ref.~\cite{Radyushkin:2015gpa}, by including the one-gluon exchange contribution, which is encoded by the perturbative evolution kernel,  to the present result, one can explore the perturbative evolution of QDA as well. The related works are under progress and appears elsewhere.
\section*{Acknowledgements}
The author is grateful to H.~D.~Son (Bochum) for fruitful discussions. The present work was partially supported by the research fund of Pukyong National University (PKNU).


\begin{thebibliography}{99}
\bibitem{Collins:1989gx}
  J.~C.~Collins, D.~E.~Soper and G.~F.~Sterman,
  Adv.\ Ser.\ Direct.\ High Energy Phys.\  {\bf 5}, 1 (1989).
\bibitem{Lepage:1979za}
  G.~P.~Lepage and S.~J.~Brodsky,
  Phys.\ Rev.\ Lett.\  {\bf 43}, 545 (1979)
  Erratum: [Phys.\ Rev.\ Lett.\  {\bf 43}, 1625 (1979)].
\bibitem{Lepage:1979zb}
  G.~P.~Lepage and S.~J.~Brodsky,
  Phys.\ Lett.\  {\bf 87B}, 359 (1979).
\bibitem{Efremov:1979qk}
  A.~V.~Efremov and A.~V.~Radyushkin,
  Phys.\ Lett.\  {\bf 94B}, 245 (1980).
\bibitem{Lepage:1980fj}
  G.~P.~Lepage and S.~J.~Brodsky,
  Phys.\ Rev.\ D {\bf 22}, 2157 (1980).
\bibitem{Chernyak:1983ej}
  V.~L.~Chernyak and A.~R.~Zhitnitsky,
  Phys.\ Rept.\  {\bf 112}, 173 (1984).
\bibitem{Behrend:1990sr}
  H.~J.~Behrend {\it et al.} [CELLO Collaboration],
  Z.\ Phys.\ C {\bf 49}, 401 (1991).
\bibitem{Gronberg:1997fj}
  J.~Gronberg {\it et al.} [CLEO Collaboration],
  Phys.\ Rev.\ D {\bf 57}, 33 (1998).
\bibitem{Aubert:2009mc}
  B.~Aubert {\it et al.} [BaBar Collaboration],
  Phys.\ Rev.\ D {\bf 80}, 052002 (2009).
\bibitem{Uehara:2012ag}
  S.~Uehara {\it et al.} [Belle Collaboration],
  Phys.\ Rev.\ D {\bf 86}, 092007 (2012).
\bibitem{Schmedding:1999ap}
  A.~Schmedding and O.~I.~Yakovlev,
  Phys.\ Rev.\ D {\bf 62}, 116002 (2000).
\bibitem{Chernyak:1977fk}
  V.~L.~Chernyak, A.~R.~Zhitnitsky and V.~G.~Serbo,
  JETP Lett.\  {\bf 26}, 594 (1977)
  [Pisma Zh.\ Eksp.\ Teor.\ Fiz.\  {\bf 26}, 760 (1977)].
\bibitem{Bakulev:2002uc}
  A.~P.~Bakulev, S.~V.~Mikhailov and N.~G.~Stefanis,
  Phys.\ Rev.\ D {\bf 67}, 074012 (2003).
\bibitem{Bakulev:2003cs}
  A.~P.~Bakulev, S.~V.~Mikhailov and N.~G.~Stefanis,
  Phys.\ Lett.\ B {\bf 578}, 91 (2004).
\bibitem{Braun:1988qv}
  V.~M.~Braun and I.~E.~Filyanov,
  Z.\ Phys.\ C {\bf 44}, 157 (1989).
\bibitem{Bakulev:1991ps}
  A.~P.~Bakulev and A.~V.~Radyushkin,
  Phys.\ Lett.\ B {\bf 271}, 223 (1991).
\bibitem{Bakulev:1994su}
  A.~P.~Bakulev and S.~V.~Mikhailov,
  Z.\ Phys.\ C {\bf 68}, 451 (1995).
\bibitem{Bakulev:2001pa}
  A.~P.~Bakulev {\it et al.},
  Phys.\ Lett.\ B {\bf 508}, 279 (2001)
  Erratum: [Phys.\ Lett.\ B {\bf 590}, 309 (2004)].
\bibitem{Bakulev:2005vw}
  A.~P.~Bakulev and N.~G.~Stefanis,
  Nucl.\ Phys.\ B {\bf 721}, 50 (2005).
\bibitem{Dalley:2002nj}
  S.~Dalley and B.~van de Sande,
  Phys.\ Rev.\ D {\bf 67}, 114507 (2003).
\bibitem{Petrov:1998kg}
  V.~Y.~Petrov, M.~V.~Polyakov, R.~Ruskov, C.~Weiss and K.~Goeke,
  Phys.\ Rev.\ D {\bf 59}, 114018 (1999).
\bibitem{Nam:2006au}
  S.~i.~Nam, H.~-Ch.~Kim, A.~Hosaka and M.~M.~Musakhanov,
  Phys.\ Rev.\ D {\bf 74}, 014019 (2006).
\bibitem{Nam:2006sx}
  S.~i.~Nam and H.~-Ch.~Kim,
  Phys.\ Rev.\ D {\bf 74}, 076005 (2006).
\bibitem{Praszalowicz:2001wy}
  M.~Praszalowicz and A.~Rostworowski,
  Phys.\ Rev.\ D {\bf 64}, 074003 (2001).
\bibitem{Praszalowicz:2001pi}
  M.~Praszalowicz and A.~Rostworowski,
  Phys.\ Rev.\ D {\bf 66}, 054002 (2002).
\bibitem{RuizArriola:2002bp}
  E.~Ruiz Arriola and W.~Broniowski,
  Phys.\ Rev.\ D {\bf 66}, 094016 (2002).
\bibitem{Ji:2013dva}
  X.~Ji,
  Phys.\ Rev.\ Lett.\  {\bf 110}, 262002 (2013).
\bibitem{Zhang:2017bzy} 
  J.~H.~Zhang, J.~W.~Chen, X.~Ji, L.~Jin and H.~W.~Lin,
  Phys.\ Rev.\ D {\bf 95}, no. 9, 094514 (2017).
\bibitem{Carlson:2017gpk} 
  C.~E.~Carlson and M.~Freid,
  Phys.\ Rev.\ D {\bf 95}, no. 9, 094504 (2017).
\bibitem{Radyushkin:2017gjd} 
  A.~V.~Radyushkin,
  Phys.\ Rev.\ D {\bf 95}, no. 5, 056020 (2017).
\bibitem{Radyushkin:2015gpa}
  A.~V.~Radyushkin,
  Phys.\ Rev.\ D {\bf 93}, no. 5, 056002 (2016).
\bibitem{Diakonov:2002fq}
  D.~Diakonov,
  Prog.\ Part.\ Nucl.\ Phys.\  {\bf 51}, 173 (2003).
\bibitem{Diakonov:1985eg}
  D.~Diakonov and V.~Y.~Petrov,
  Nucl.\ Phys.\ B {\bf 272}, 457 (1986).
\bibitem{Diakonov:1995qy}
  D.~Diakonov, M.~V.~Polyakov and C.~Weiss,
  Nucl.\ Phys.\ B {\bf 461}, 539 (1996).
\bibitem{Nam:2006ng}
  S.~i.~Nam and H.~-Ch.~Kim,
  Phys.\ Lett.\ B {\bf 647}, 145 (2007).
\bibitem{Nam:2007gf}
  S.~i.~Nam and H.~-Ch.~Kim,
  Phys.\ Rev.\ D {\bf 77}, 094014 (2008).
\bibitem{RuizArriola:2006jge}
  E.~Ruiz Arriola and W.~Broniowski,
  Phys.\ Rev.\ D {\bf 74}, 034008 (2006).
\bibitem{Oganesian:2015ucv}
  A.~G.~Oganesian {\it et al},
  Phys.\ Rev.\ D {\bf 93}, no. 5, 054040 (2016).
\end{thebibliography}
\end{document}